\documentclass[prl, aps, twocolumn, amssymb, superscriptaddress, 10pt]{revtex4-2}
\usepackage{amssymb}
\usepackage{amsbsy}
\usepackage{amsmath}
\usepackage{amsthm}
\usepackage{graphicx}
\usepackage{graphics}
\usepackage{setspace}
\usepackage{array}
\usepackage{color}
\usepackage{xcolor}
\usepackage{fontenc}
\usepackage{textcomp}
\usepackage{bm}
\usepackage{pifont}
\usepackage[bookmarks=false,linkcolor=blue,urlcolor=blue,colorlinks,citecolor=blue]{hyperref}
\usepackage{soul}
\usepackage[T1]{fontenc}
\usepackage{mathdots}

\usepackage{dsfont}
\usepackage{bbold}

\DeclareMathOperator{\Tr}{Tr}
\newcommand{\var}{\text{Var}}

\newcommand{\n}{\nonumber}

\newcommand{\blue}[1]{{\color{blue}{#1}}}

\newcommand{\bsub}{\begin{subequations}}
\newcommand{\esub}{\end{subequations}}

\definecolor{darkred}{rgb}{0.8,0,0}
\definecolor{royalblue}{rgb}{0.0, 0.14, 0.4}
\definecolor{magenta}{cmyk}{0,.9,0,0.2}
\definecolor{amethyst}{rgb}{0.6, 0.4, 0.8}

\newcommand{\beginsupplement}{
        \setcounter{table}{0}
        \renewcommand{\thetable}{S\arabic{table}}
        \setcounter{figure}{0}
        \renewcommand{\thefigure}{S\arabic{figure}}
        \setcounter{equation}{0}
        \renewcommand{\theequation}{S\arabic{equation}}
        \setcounter{section}{0}
        \renewcommand{\thesection}{\Alph{section}}
        \setcounter{subsection}{0}
        \renewcommand{\thesubsection}{\arabic{subsection}}
}

\begin{document}
\title{Local and Non-local Entanglement Witnesses of Fermi Liquid}

\author{Yiming Wang}
\thanks{These authors contributed equally.}
\affiliation{Department of Physics \& Astronomy,  Extreme Quantum Materials Alliance, Smalley-Curl Institute, Rice University, Houston, Texas 77005, USA}

\author{Yuan Fang}
\thanks{These authors contributed equally.}
\affiliation{Department of Physics \& Astronomy,  Extreme Quantum Materials Alliance, Smalley-Curl Institute,
Rice University, Houston, Texas 77005, USA}

\author{Fang Xie}
\thanks{These authors contributed equally.}
\affiliation{Department of Physics \& Astronomy,  Extreme Quantum Materials Alliance, Smalley-Curl Institute, Rice University, Houston, Texas 77005, USA}
\affiliation{Rice Academy of Fellows, Rice University, Houston, Texas 77005, USA}

\author{Qimiao Si}
\affiliation{Department of Physics \& Astronomy,  Extreme Quantum Materials Alliance, Smalley-Curl Institute,
Rice University, Houston, Texas 77005, USA}

\begin{abstract}
    There is a growing interest both in utilizing entanglement means to characterize many-body systems and in uncovering their entanglement depth. Motivated by recent findings that the spin quantum Fisher information witnesses amplified multipartite entanglement of strange metals and characterizes their loss of quasiparticles, we study the quantum Fisher information in various cases of Fermi liquid. We show that local operators generically do not witness any multipartite entanglement in a Fermi liquid, but non-local many-body operators do. Our results point to novel experimental means to detect the entanglement depth of metallic fermionic systems and, in general, open a new avenue to the emerging exploration of entanglement in quantum materials.
\end{abstract}

\maketitle

\blue{\emph{Introduction}}---
Entanglement describes the entwining of particles such that their quantum states cannot be separately described~\cite{bell2004speakable,zeng2019quantum}. 
It is a key resource in quantum information processing and quantum computation~\cite{nielsen2001quantum,Chitambar2019Quantum,Tan2021Fisher}. 
In many-body systems, entanglement is usually characterized by such quantities as entanglement entropy
~\cite{amico2008entanglement,kitaev2006topological,li2008entanglement,Wolf2008Area}.
However, these quantities are hard to experimentally measure.
They also depend on the choice of bi-partition of the Hilbert space and do not capture any multi-partite entanglement. 
Recently, quantum Fisher information (QFI) has been recognized as an entanglement witness of many-body systems~\cite{hyllus2012fisher,Toth2012Multipartite,hauke2016measuring,Liu2020Quantum,Scheie2024Tutorial}. 
QFI is a function of both the density matrix and the summation of some local and mutually commuting hermitian operators.
These operators define a multi-partition of the Hilbert space.
The QFI detects the entanglement among the partitions, providing a lower bound of the entanglement depth of a quantum state, i.e., the minimal number of partitions that the state is entangled across~\cite{hyllus2012fisher}. 
One important advantage of the QFI is that it can be determined from the correlation functions of the local operators, and thus is experimentally measurable~\cite{hauke2016measuring}. 

Strongly correlated systems can be especially collective, as exemplified by the quantum spin liquids and  fractional quantum Hall states, which are expected to be strongly entangled~\cite{wen2004quantum,kitaev2006topological,li2008entanglement}. 
Strange metals represent another example in this general category.
It was recently shown that the QFI of the local spin operators witnesses multipartite entanglement in the strange metals \cite{Fang2024Amplified,Mazza2024Quantum}, providing a new characterization of the loss of Landau quasiparticles \cite{hu2022quantumc,Si2001,Colemanetal,senthil2004a,Liyang-Chen2023,Pfau-2012}.

This development raises a basic question -- what happens to multipartite entanglement in Fermi liquids? Metallic fermionic systems with a Fermi surface is expected to have a large amount of entanglement; for example, calculations have shown that their entanglement entropy violates the area law~\cite{Wolf2006Violation}. 
Especially from the perspective of revealing the entanglement depth, it would be important to address whether there are measurable observables that can probe any aspect of the entanglement in such systems.

In this work, we address this question in several examples of Fermi liquids.
We show that local operators generically do not witness multipartite entanglement in Fermi liquids, but non-local operators 
such as those constructed by a Jordan-Wigner transformation can.
Our results have important implications for experiments in a wide variety of gapless fermionic systems.

\blue{\emph{Quantum Fisher information as entanglement witness of fermionic systems}}---
For a pure state $|\psi\rangle$, QFI of the operator 
$\widehat{\cal O}= \sum_{i=1}^{N_L} e^{i\chi_i} \widehat{\cal O}_{i}$ is the equal time correlation function~\cite{hyllus2012fisher}: 
\begin{equation}
    F_Q(|\psi\rangle, \widehat{\mathcal O}) = 4 \left( \langle \psi| \widehat{\mathcal O}^\dagger \widehat{\mathcal O} |\psi\rangle - |\langle \psi| \widehat{\mathcal O} |\psi \rangle|^2 \right) \,.
    \label{eq:qfi-variance}
\end{equation}
Here, $\chi_i \in \mathbb{R}$ is a position dependent phase factor and $\widehat{\cal O}_{i}$ is a local Hermitian operator that acts on the $i$-th site among the $N_{L}$ sites (modes).
For further discussion on the effect of this phase factor, see Sec.~A in the Supplemental Materials (SM)~\cite{sm}.
At finite temperature $T$, the QFI can be expressed in terms of the dynamical susceptibility~\cite{hauke2016measuring} as follows:
\begin{equation}
\label{eqn:QFI_sus}
    F_{Q} = \frac{4}{\pi} \int_{0}^\infty \tanh{\frac{ \hbar \omega}{2 k_B T}} \chi''_{\widehat{\cal O}}(\omega) d\omega \, .
\end{equation}
Here $\chi''_{\widehat{\cal O}}(\omega)$ is the imaginary part of the susceptibility $\chi_{\widehat{\cal O}}(\omega)= i\int^\infty_0 dt\, e^{i\omega t} \Tr\left( \rho [{\widehat{\cal O}}^\dagger(t),{\widehat{\cal O}}(0)] \right)$.

Akin to the Bell's inequality for the case of a few particles, here the QFI can be used to detect the entanglement of a quantum system via a bound on the entanglement depth.
Consider a system with $N_{L}$ sites (modes), and the local operator $\widehat{\cal O}_i$ acting on the $i$-th site. 
The QFI density $f_Q \equiv F_{Q}/{N_L}$ of an $m$-partite entangled state is bounded by~\cite{hyllus2012fisher,Fang2024Amplified,Pezze2018Quantum}:
\begin{equation}
    \label{eqn:bound}
    f_Q \leq m (h_{\text{max}}-h_{\text{min}})^2 \, ,
\end{equation}
where $h_{\text{max}}$ and $h_{\text{min}}$ are the maximum and minimum eigenvalues of the local operator $\widehat{\cal O}_i$.
Accordingly, one is interested in the normalized QFI density:
\begin{align}
   \text{nQFI} = f_Q/(h_{\text{max}}-h_{\text{min}})^2 \,. 
\end{align}
If the nQFI is larger than $m$, the system is at least $(m+1)$-partite entangled, i.e., the entanglement depth is at least $m+1$. 
Most of the studies on the entanglement witnesses have been on quantum many spin systems, including the Greenberger-Horne-Zeilinger (GHZ) state~\cite{hyllus2012fisher}, the transverse field Ising model (TFIM)~\cite{hauke2016measuring}, and other spin-based models and materials 
~\cite{Scheie2021Witnessing,*Scheie2023Erratum,scheie2023proximate,Laurell2021Quantifying,Laurell2022Magnetic,hales2023witnessing,Lambert2020Revealing,George2020Experimental,Pratt2022Spin}. 
Except for a few isolated cases~\cite{Baykusheva2023Witnessing,Fang2024Amplified}, the issue of entanglement witnesses of Fermi liquids has not yet been addressed.

Using the QFI as an entanglement witness only requires that the ``local'' operators commute $[\widehat{\cal O}_i, \widehat{\cal O}_j]=0$. In other words, each operator $\widehat{\cal O}_i$ is a bounded Hermitian operator acting on the local Hilbert subspace ${\cal H}_i$.
This defines a multi-partite partition of the Hilbert space, and the QFI can be used to detect the entanglement among the partitions.

\begin{figure}[t]
    \centering
    \includegraphics[width=\linewidth]{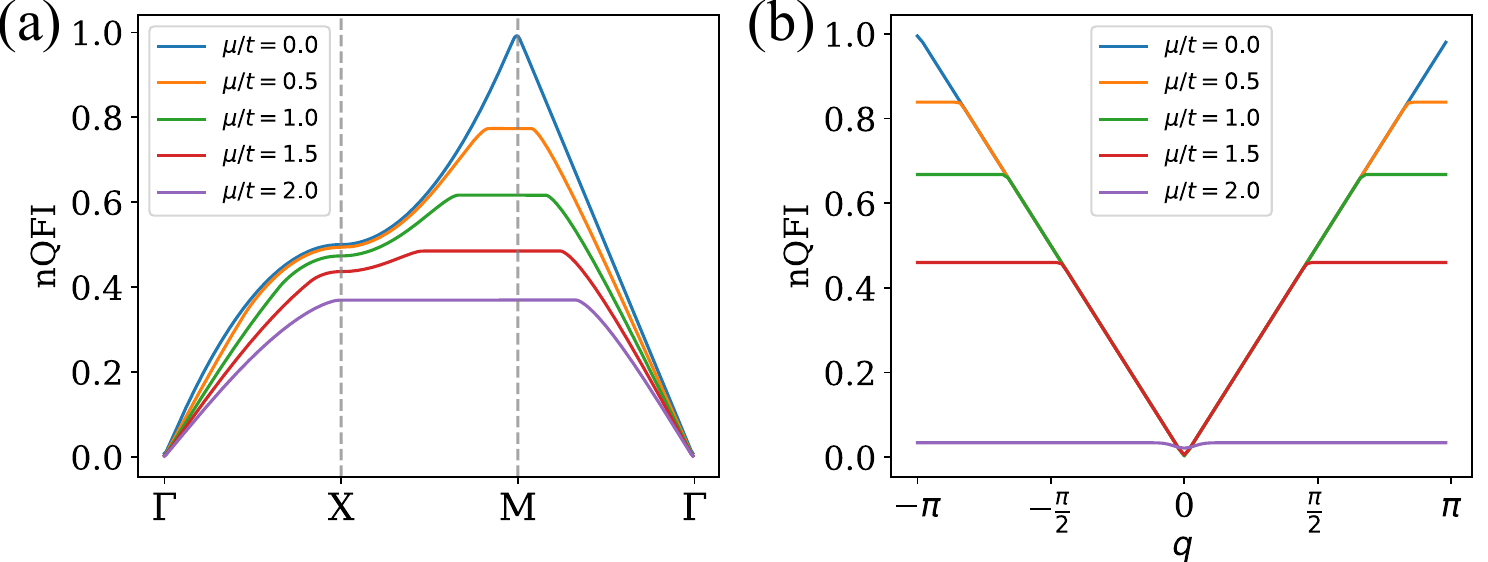}
    \caption{The 
    nQFI of the local spin operator in a noninteracting metal for (a) the 2D square lattice with nearest-neighbor hopping, and (b) the 1D chain with nearest-neighbor hopping (Eq.~(\ref{eqn:H0})). Varying chemical potentials are considered, ranging from the half filling $\mu/t=0$ to values away from the half filling $\mu/t=0.5,1.0,1.5,2.0$.
    } 
    \label{fig:localslater}
\end{figure}

\blue{\emph{Quantum Fisher information of local operators for noninteracting electrons with a Fermi surface}}---
We start from a noninteracting electron system that has a Fermi surface, in any dimension, with the Hamiltonian 
\begin{equation}
h_0=\sum_{\bm{k},\sigma}\epsilon_{\bm{k}}c^{\dagger}_{\bm{k}\sigma}c_{\bm{k}\sigma} \,,
\end{equation}
where $\epsilon_{\bm{k}}$ denotes the band dispersion, $\bm{k}$ represents the wavevector and $\sigma= \uparrow,\downarrow$ is the spin index. 

For definiteness, we consider spin operators: $\widehat{\cal O}(\bm{q})=\sum_{i}e^{i\bm{q}\cdot \bm{r}_i}\widehat{\cal O}_{i}$ at wave vector $\bm{q}$, with $\widehat{\cal O}_{i}=s^{z}_{i}=\frac{1}{2}\sum_{\sigma\sigma'}c^{\dagger}_{i\sigma}[\sigma_{z}]_{\sigma\sigma'}c_{i\sigma'}$.
Following Eq.~(\ref{eq:qfi-variance}), the QFI density has the following expression:
\begin{align}
    f^{0}_Q(\bm q)=&\frac{4}{N_L}\sum_{i,j}e^{i\bm{q}\cdot(\bm{r}_i-\bm{r}_j)}\langle  \nonumber s^{z}_{i}s^{z}_{j}\rangle\\
    =&\frac{2}{N_L}\sum_{\bm{k}}f(\epsilon_{\bm k})(1-f(\epsilon_{\bm k+ \bm q}))\,,
\end{align}
where $f(\epsilon_{\bm k})=\theta(-\epsilon_{\bm k})$ is the Fermi-Dirac distribution function at zero temperature.
Its upper bound is determined from the arithmetic mean-geometric mean inequality:
\begin{align}\label{eqn:bound_slater}
f_Q^{0}(\bm q) &\leq \frac{1}{N_L}\sum_{\bm k}(f(\epsilon_{\bm k})^2 + (1-f(\epsilon_{\bm k+\bm q}))^{2})\nonumber
\\ &= \frac{1}{N_L}\sum_{\bm k}(f(\epsilon_{\bm k}) + 1-f(\epsilon_{\bm k+\bm q})) \nonumber \\
&= 1\,.
\end{align}
Here we have set $T=0$.
The equality holds when and only when ${\rm sgn}(\epsilon_{\bm k}) = -{\rm sgn}(\epsilon_{\bm k+\bm q})$ for every single $\bm k$ (an equality that is stronger than the usual Fermi-surface nesting condition).
In the case of electrons on a square lattice in two dimensions (2D), this condition is satisfied when $\bm{q}=(\pi,\pi)$ at half-filling, where infinitesimal Hubbard interactions can induce a spin-density wave instability. We note that this upper bound is valid for any noninteracting electron system.

To further discuss the above bound, we have calculated the nQFI at different wave vectors ($\bm{q}$) and different fillings of noninteracting electron systems, for the case of a 1D chain and a 2D square lattice.
The Hamiltonian is 
\begin{equation} \label{eqn:H0}
    H_0 = -t \sum_{\langle i,j \rangle, \sigma} c_{i\sigma}^\dagger c_{j\sigma} - \mu \sum_{i, \sigma} c_{i\sigma}^\dagger c_{i\sigma} \,,
\end{equation}
where $t$ is the amplitude for the nearest-neighbor hopping and $\mu$ is the chemical potential. 
The results are shown in Fig.\,\ref{fig:localslater}(a) for 1D and Fig.\,\ref{fig:localslater}(b) for 2D. The vanishing QFI at $\bm{q}=0$ implies the total spin conservation: $[\widehat{\cal O}(0),H]=0$, such that $\langle \widehat{\cal O}(0)^{2}\rangle= \langle \widehat{\cal O}(0)\rangle^{2}$ at zero temperature.

\begin{figure}[t]
    \centering
    \includegraphics[width=\linewidth]{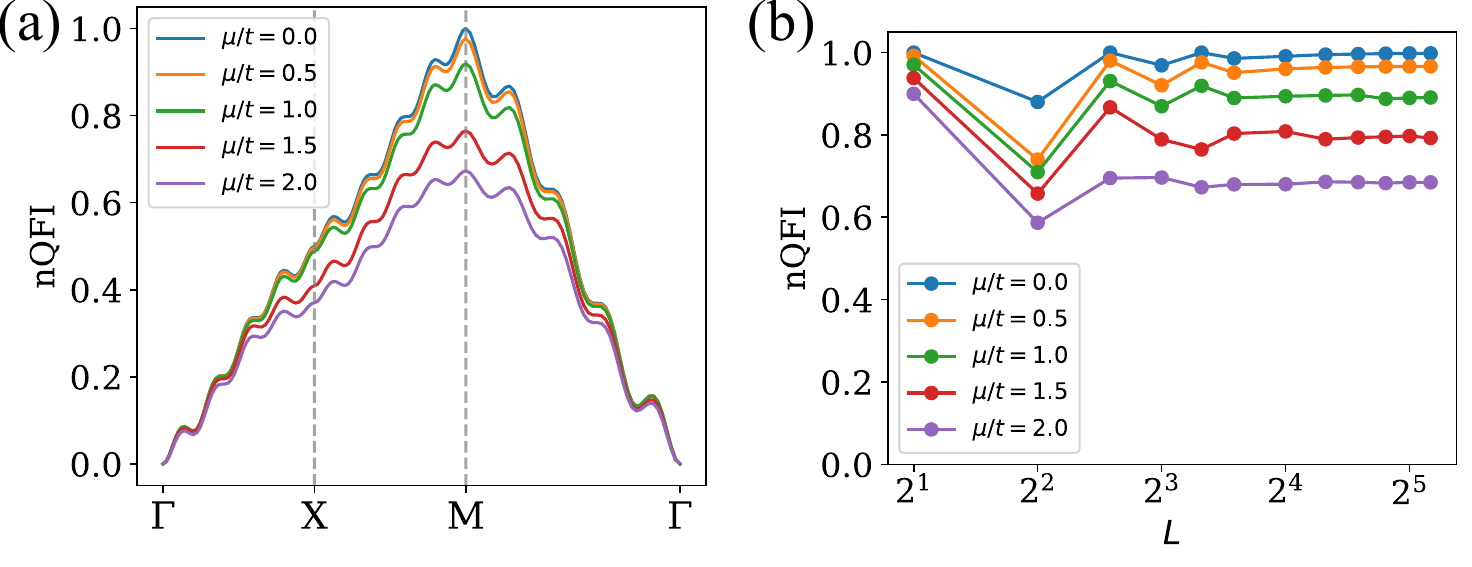}
    \caption{(a) The nQFI of the local spin operator for the 2D $d$-wave superconductor (Eq.~(\ref{eqn:HbdG})) with $\Delta/t=1$, the system size $10\times 10$ and varying chemical potential.
    (b) The nQFI at ${\bm q}=(\pi,\pi)$ v.s. the system size $L\times L$  of the local spin operator for the 2D $d$-wave superconductor at varying chemical potentials.
    } 
    \label{fig:localbdg}
\end{figure}

When the spin $SU(2)$ symmetry is present, the nQFI of the local spin operator is the same as the nQFI of the local charge operator (see Sec.~B in SM~\cite{sm}). 
Therefore, the upper bound we proved for the local spin nQFI also applies to its charge counterpart.

Our results extend naturally to multiorbital systems. We can define the generalized density operators
$\widehat{\cal O}(\bm{q})=\sum_{i\alpha}e^{i\bm{q}\cdot \bm{r}_i}\widehat{\cal O}_{i\alpha}$ at wave vector $\bm{q}$, with $\widehat{\cal O}_{i\alpha}=\frac{1}{2}\sum_{\sigma\sigma'}c^{\dagger}_{i\alpha\sigma}[\sigma^{a}]_{\sigma\sigma'}c_{i\alpha\sigma'}$ for each orbital $\alpha$ and each site $i$, and $a = 0$ or $z$ represents the charge or spin channel. 
Despite the complexity introduced by the multiplicity of the orbitals and diverse band structures, the nQFI remains universally bounded by 1 for both the charge and spin channels (see Sec.~C in SM~\cite{sm}).

As comparison and for completeness, we also study the superconducting states in a $d$-wave superconductor in the 2D case.
The Hamiltonian, defined on the square lattice, is
\begin{equation}\label{eqn:HbdG}
    H_{\rm BdG} = H_0
+ \sum_{\langle i,j \rangle} ( \Delta_{ij} c_{i\uparrow}^\dagger c_{j\downarrow}^\dagger + h.c. ) \,,
\end{equation}
where $\Delta_{ij}$ is equal to $\Delta$ for pairing on the nearest-neighbor bond along the $x$ direction and $-\Delta$ for its $y$-direction counterpart.
For illustration, we choose $\Delta/t=1$ for the calculation.
Consider the spin operator $\widehat{\cal O}({\bm q})=\sum_i e^{i {\bm q} \cdot {\bm r}_i} s^z_i$.
The equal-time correlator is straightforwardly calculated using Wick's theorem~\cite{sm}.
Fig.\,\ref{fig:localbdg}~(a) shows the spin nQFI vs. $\bm q$ for a system size of $L \times L$ at $L=10$.
The size dependence at ${\bm q}=(\pi,\pi)$ is displayed in Fig.\,\ref{fig:localbdg}~(b).
The local spin nQFI is seen not to exceed $1$.
A general discussion in Sec.~D of the SM~\cite{sm} shows that the local spin nQFI for any Bogoliubov-de Gennes (BdG) state is upper bounded by $1$.
Thus, we conclude that the local spin nQFI does not witness any multipartite entanglement for the BdG states as well.

\begin{figure}[t]
    \centering
        \centering
        \includegraphics[width=\linewidth]{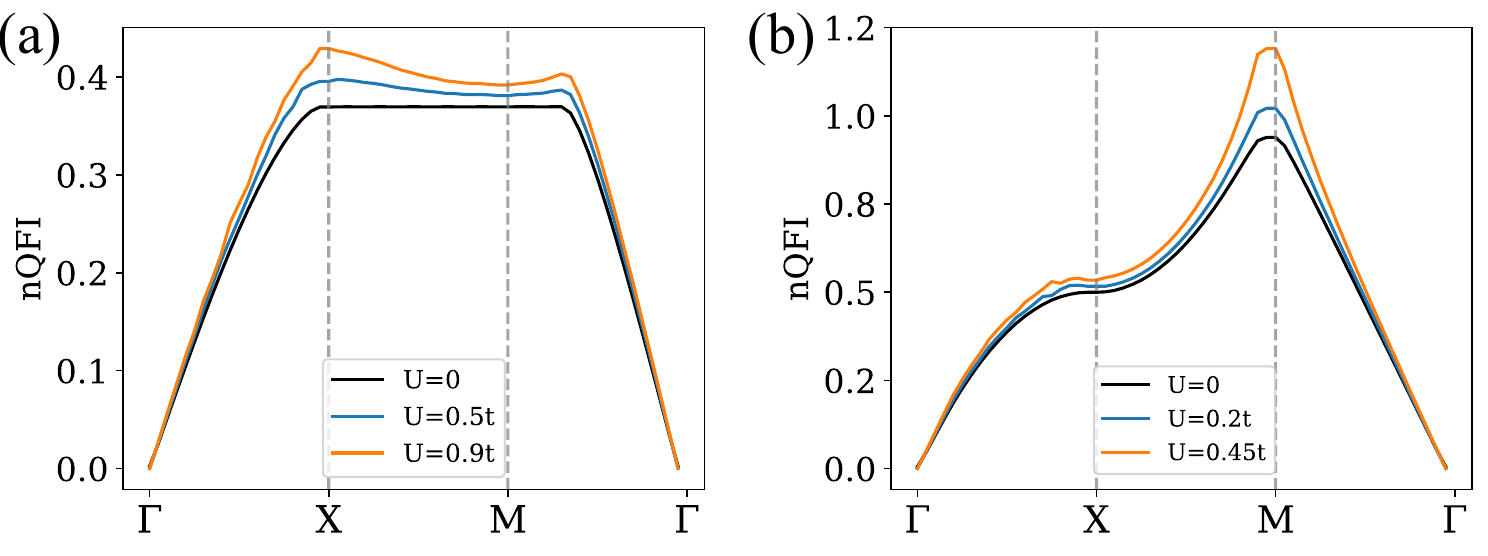}
    \caption{nQFI of the 2D interacting Hubbard model on a square lattice with nearest-neighbor hopping (Eq.~(\ref{eqn:HRPA})) using an RPA treatment at (a) \( \mu=2t \) for \( U=0, 0.5t, 0.9t \) and (b) \( \mu=0.1t \) for \( U=0, 0.2t, 0.45t \). The instability to spin density wave (SDW) order occurs at (a) \( U_c=0.91t \) and (b) \( U_c=0.48t \).}
    \label{fig:PRA}
\end{figure}

\blue{\emph{Quantum Fisher information of local operators in a Fermi liquid}}---
The upper bounds imply that the local spin operators cannot serve as entanglement witnesses in these noninteracting electron systems. 
The presence of electronic correlations can, in principle, break this bound. To see this, we perform a random phase approximation (RPA) analysis of the Hubbard model: 
\begin{equation}\label{eqn:HRPA}
    H = H_0 + U\sum_{i}n_{i\uparrow}n_{i\downarrow}\,.
\end{equation}
The dynamical spin susceptibility at the RPA level has the following form:
\begin{align}
    \chi^{\prime\prime}_{RPA}(\bm{q},\omega)=\frac{\chi^{\prime\prime}_{0}(\bm{q},\omega)}{[1-U\,\chi_{0}^{\prime}(\bm{q},\omega)]^2+[U\chi^{\prime\prime}_{0}(\bm{q},\omega)]^2} \, ,
\end{align}
where $\chi_{0}^{\prime}(\bm{q},\omega)$ and $\chi^{\prime\prime}_{0}(\omega,\bm{q})$ are the real and imaginary parts of the spin susceptibility of the noninteracting electrons. The QFI of the interacting systems can then be determined through Eq.~(\ref{eqn:QFI_sus}), with $\chi^{\prime\prime}_{RPA}(\bm{q},\omega) = \chi^{\prime\prime}_{\widehat{\cal O}}(\omega)/N_{L}$.
In general, the QFI for the repulsive interaction case at the RPA level exceeds that of its noninteracting electron counterpart (see Sec.~III in the SM~\cite{sm}): 
\begin{align}
    f_Q^{RPA}(\bm{q})\geq f_Q^{0}(\bm{q})\,,
\end{align}
given that repulsive interactions enhances the dynamical spin susceptibility: $\chi^{\prime\prime}_{RPA}(\omega,\bm{q})\geq \chi^{\prime\prime}_{0}(\omega,\bm{q})$. 
The nQFI for the interacting electrons with different values of the Hubbard interaction calculated at the RPA level is shown in Fig.\,\ref{fig:PRA}(a)(b). In (a), we consider $\mu=2t$, where the instability to the spin density wave order occurs at $U_{c}=0.91t$. The results indicate that while it is enhanced by the interactions, the nQFI remains below 1, implying that no entanglement is detected. 
In some fine tuned region close to the half filling, the nQFI can slightly exceed $1$ (witnessing bipartite entanglement), as illustrated in (b) for the particular case close to the half-filling at $\mu= 0.1t$.

\blue{\emph{Quantum Fisher information of non-local operators for fermion systems}}---
The above results imply that local operators are not adequate to fully capture the entanglement structure of the fermion systems with a Fermi surface.
Recognizing that metallic fermion systems with a Fermi surface have a large entanglement entropy, we ask whether there are other operators that can serve as entanglement witnesses.
To this end, we consider non-local many-body bosonic operators.
Our strategy is to attach non-local string operators to each fermionic operator via the Jordan-Wigner transformation.

We consider spinless noninteracting electrons in (1) a 1D chain and (2) a ladder of two coupled chains, both with a nearest-neighbor hopping.
The Hamiltonians are
\begin{align}
    H_{1} &=  -t \sum_{\langle i,j \rangle, } c_{i}^\dagger c_{j} - \mu \sum_{i, } c_{i}^\dagger c_{i} \,, \label{eqn:H1}\\
    H_{2} &= H_1^A+H_1^B -t \sum_i (c^\dagger_{i,A} c_{i,B} + c^\dagger_{i,B} c_{i,A} ) \,, \label{eqn:H2}
\end{align}
where $A,B$ label the two chains.
$H_{1}$ is the spinless version of $H_0$ (Eq.~(\ref{eqn:H0})).

For the 1D chain, the Jordan-Wigner transformation maps fermionic operators $c^\dagger_i$, $c_i$ to hard-core bosonic operators via a flux attaching term in 1D~\cite{Jordan1993}:
\begin{align}
    \label{eqn:JW}
    a^\dagger_i =  c_i^\dagger {\hat \Theta_i}^{-1} \,, \quad a_i = {\hat \Theta_i} c_i\,,
\end{align}
where $\hat \Theta_i=e^{-i\pi \sum_{j<i}c^\dagger_jc_j}$ in 1D.
The operators $a_i$ and $a_i^\dagger$ satisfy the commutation relations:
\begin{align}
    \label{eqn:commutation}
    &[a_i, a_j] = [a_i^\dagger, a_j^\dagger] = 0 \,, ~~ \forall i,j \,, \notag \\
    &[a_i, a_j^\dagger] = 0, ~ \forall i\neq j, \quad \{a_i, a_i^\dagger\}=1 \,.
\end{align}
These operators describe hard-core bosons and do not change the Hilbert space of the system.

In the $L\times 2$ ladder system, the Jordan-Wigner transformation can be defined by the extended string operators $\hat \Theta_i=e^{-i\theta^{\text{sgn}}_{ij}\sum_{j\neq i}c^\dagger_jc_j}$ where $\exp(-i\theta^{\text{sgn}}_{ij}) = \text{sgn}(x_i-x_j) + \text{sgn}(y_i-y_j) \delta_{x_i,x_j}$, $x_i=1,\dots,L$ and $y_i=A,B$ are the $x$ coordinate and ladder index of the $i$-th mode. Here $\text{sgn}(0)=0$ is imposed for this definition.
This amounts to a special form of the 2D Jordan Wigner transformation~\cite{Fradkin1989Jordan,Batista2001Generalized,ELIEZER1992118}.

These hard-core bosonic operators can be used for considering the QFI of the underlying fermionic systems. 
Although non-local in real space and involving many fermion operators, these commuting operators define disconnected Hilbert subspaces and fulfill the criteria of using QFI as entanglement witness.
Specifically, we consider $\widehat{\cal O}({\bm q}) = \sum_i e^{i{\bm q}\cdot {\bm r}_i}\widehat{\cal O}_i$ in which
\begin{equation}
    \widehat{\cal O}_i = a_i + a_i^\dagger \quad {\rm or} \quad \widehat{\cal O}_i =  i(a_i - a_i^\dagger)
\end{equation}
as the local operator in the transformed basis.
Its normalized QFI density is $\text{nQFI}=f_Q/4$, since the local Hermitian operator has eigenvalues $\pm 1$~\cite{sm}.

\begin{figure}[t]
    \centering
    \includegraphics[width=\linewidth]{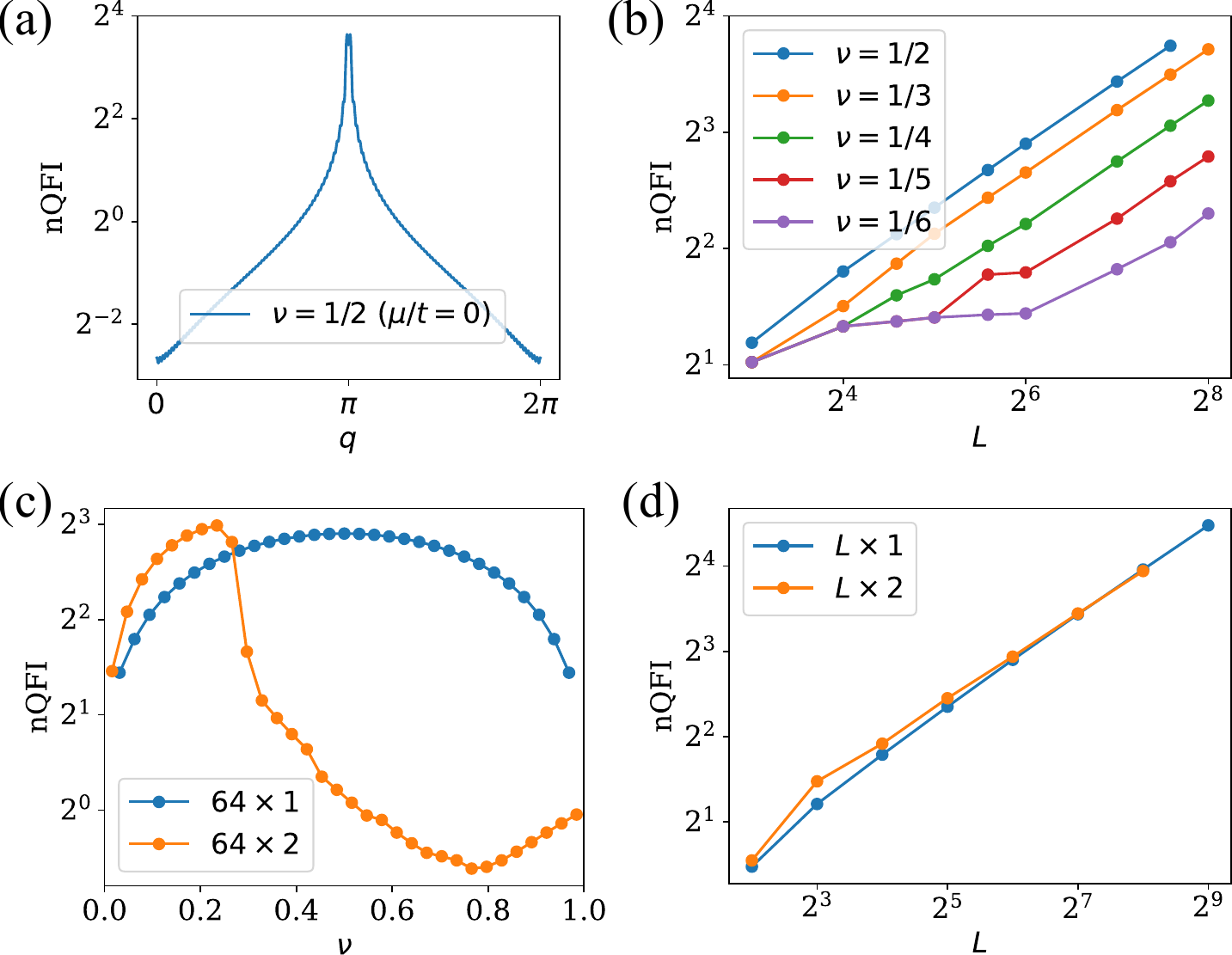}
    \caption{(a) 
    nQFI of 1D 
    noninteracting electrons (Eq.~(\ref{eqn:H1}))
    at half filling $\mu=0$ with $L=24$.
    (b) nQFI of 1D noninteracting electrons at $q=\pi$
    v.s. the system size $L$ at different fillings $\nu$.
    (c) 
    nQFI of non-local operator for noninteracting electrons on 1D chain and two-leg ladder (Eq.~(\ref{eqn:H2})) at different fillings and ${\bm q}=(\pi,\pi)$.
    (d) nQFI of non-local operator for noninteracting electrons on 1D chain and two-leg ladder at ${\bm q}=(\pi,\pi)$ with different system size $L$.
    } 
    \label{fig:nonlocal}
\end{figure}

We apply the framework introduced above for the two cases. 
Here we focus on the operator $\widehat{\cal O}({\bm q}) = \sum_i e^{i{\bm q}\cdot {\bm r}_i}(a_i + a_i^\dagger)$.
The involved correlation functions of this string operator can be calculated by evaluating a Toeplitz matrix~\cite{PFEUTY197079,Barouch1971Statistical,sachdev1999quantum,Mbeng2024Ising} (see the SM~\cite{sm}).

We first consider the 1D noninteracting metal, with the Hamiltonian $H_1$ defined above.
Fig.\,\ref{fig:nonlocal}(a) presents the nQFI of the Jordan-Wigner transformation operators with $\theta^{\text{sgn}}_{ij}$ for the half-filled case.
It shows the nQFI as peaked at $q=\pi$.
We now set the filling factor $\nu = N/N_L$ where $N$ is the number of filled states and $N_L=L\times 1$ is the number of sites, and $q=\pi$.
The nQFIs vs. the system size $L$ at different fillings are calculated and shown in Fig.\,\ref{fig:nonlocal}(b). 
The fitted slope of the half filling case within the size regime of our study is $n=0.58$, i.e., ${\rm nQFI} \propto N_L^n$.
We see that the nQFI scales with the system size, demonstrating a divergent multi-partite entanglement in the thermodynamic limit. 
Such a divergence in the thermodynamic limit at $T=0$ suggests that this non-local nQFI will be strongly temperature dependent, increasing with decreasing temperature.
Explicit calculations at nonzero temperatures will be addressed in a separate work.

Fig.\,\ref{fig:nonlocal}(c) and (d) show the nQFI of the Jordan Wigner transformed operator for noninteracting fermions of the single chain with size $L \times 1$ at $q=\pi$ and of the two chains with size $L \times 2$ at ${\bm q}=(\pi,\pi)$.
Fig.\,\ref{fig:nonlocal}(c) displays the filling dependence of the nQFI for our non-local operator, whereas 
Fig.\,\ref{fig:nonlocal}(d) shows the size dependence of this  nQFI with the particle number $N=L/2$ for both cases.

\blue{\emph{Discussion and Conclusion}}---
Several remarks are in order. First, as already mentioned, metallic free fermions with a Fermi surface have a large entanglement entropy that violates the area law~\cite{Wolf2006Violation}. 
The relation between the nQFI and the entanglement entropy is an interesting open question.
We also note that, as discussed in the contexts of resonant inelastic x-ray scattering (RIXS) and angle resolved photoemission spectroscopy (ARPES), there are other entanglement witnesses to consider for fermionic systems~\cite{malla2024detecting,Liu2024Entanglement}.

Second, the calculations for the nQFI of the non-local operators reported here are mainly for 1D systems.
Extension to the corresponding calculations of the non-local nQFI in 2D systems is nontrivial because the Jordan-Wigner transformation is not unique and the calculation of correlation functions is more involved.
We defer the discussion of these cases to a separate work.

The experimental determination of non-local quantum Fisher information of a Fermi liquid requires the measurement of correlation functions beyond the usual 2-particle level. 
For ultracold atomic fermions in optical lattices, the correlation functions of the related string operators can and have been measured~\cite{Hilker2017Revealing}. In condensed matter systems, 
techniques for measuring high-order correlation functions are being advanced in various spectroscopic contexts (eg., Ref.\,\cite{Mahmood2021Observation}).
Thus, there is prospect for measuring the non-local QFI of the Fermi liquids and witness their multipartite entanglement. Our result that the non-local nQFI is divergent in the thermodynamic limit at $T=0$ and is expected to strongly depend on temperature will facilitate its experimental measurement.

To summarize, we have considered metallic systems of noninteracting electrons with or without superconducting order parameters as well as the effect of interactions.
We show that local operators do not witness multipartite entanglement in the ground state of these Fermi liquid systems.
Using the Jordan-Wigner transformation operators as illustration, we show that many-body operators non-local in real space have normalized QFI densities that scale with the system size in certain low-dimensional cases and thus witness multipartite entanglement in the Fermi liquids. Our work raises the prospect for experimentally detecting multipartite entanglement in many-body fermionic systems.
More generally, how to probe entanglement in quantum materials has started to attract considerable interest and our work points to a new direction in this emerging field.

\blue{\emph{Acknowledgment}}---
We thank Lei Chen, Mounica Mahankali, Matteo Mitrano, Han Pu, Allen Scheie,
Ming Yi and, especially, Silke Paschen, for useful discussions. 
This work  has primarily been supported by the the NSF Grant No.\ DMR-2220603 
(local QFI, Y.W.), the AFOSR Grant No.\ FA9550-21-1-0356 (non-local QFI, Y.F., F.X.)
the Robert A. Welch Foundation Grant No.\ C-1411 and 
the Vannevar Bush Faculty Fellowship ONR-VB N00014-23-1-2870 (Q.S.). 
The majority of the computational calculations have been performed 
on the Shared University Grid at Rice funded by NSF under Grant EIA-0216467, 
a partnership between Rice University, Sun Microsystems, and Sigma Solutions, Inc., 
the Big-Data Private-Cloud Research Cyberinfrastructure
MRI-award funded by NSF under Grant No. CNS-1338099, and the Extreme Science and
Engineering Discovery Environment (XSEDE) by NSF under Grant No. DMR170109. 
The work was first presented in the eQMA workshop on Quantum Materials and Entanglement (October 2024).
Q.S. acknowledges the hospitality of the Aspen Center for Physics, 
which is supported by NSF grant No. PHY-2210452.

\bibliographystyle{apsrev4-2}
\bibliography{reference.bib}

\onecolumngrid

\setcounter{secnumdepth}{3}

\onecolumngrid
\newpage
\beginsupplement

\section*{Supplemental Materials}

\section{Quantum Fisher information}
\label{sec:QFI}
In this section we give a brief review of the quantum Fisher information (QFI) and its relation to the correlation functions in quantum systems.

For a mixed state, the density matrix can be written with the spectral decomposition $\widehat{\rho}=\sum_i \lambda_n |\lambda_n\rangle\langle\lambda_n|$. The QFI is defined as~\cite{braunstein1994statistical}:
\begin{equation}
    \label{eqn:QFI_def_par}
        F_{Q}= \sum_n \frac{(\partial_{\theta}\lambda_n)^2}{\lambda_n} + 2\sum_{nm} \frac{(\lambda_n-\lambda_m)^2}{\lambda_n+\lambda_m} \big|\langle \lambda_n| \partial_{\theta}\lambda_m \rangle \big|^2 \, .
\end{equation}
where $\theta$ is a parameter of the state.
It is a measure of the sensitivity of the state to $\theta$. The inverse of $F_Q$ gives the quantum Cram\'er-Rao bound for the estimation of $\theta$~\cite{Liu2020Quantum}.

The QFI can be used to detect  the multi-partite entanglement in a quantum system. 
Assuming the parameter in Eq.~(\ref{eqn:QFI_def_par}) is associated with a local unitary transformation $|\lambda_n\rangle \rightarrow e^{-i\theta\widehat{\cal O}}|\lambda_n\rangle$, where $\widehat{\cal O}$ is an  operator, the QFI can be written as:
\begin{equation}
    \label{eqn:QFI_def_app}
        F_{Q} = 2\sum_{nm}\frac{(\lambda_n-\lambda_m)^2}{\lambda_n+\lambda_m} {\mathcal O}_{nm} {\mathcal O}_{mn}.
\end{equation}
where ${\mathcal O}_{nm} = \langle \lambda_n| \widehat{\cal O} |\lambda_m \rangle$.

\begin{figure}
    \centering
    \includegraphics[width=0.8\linewidth]{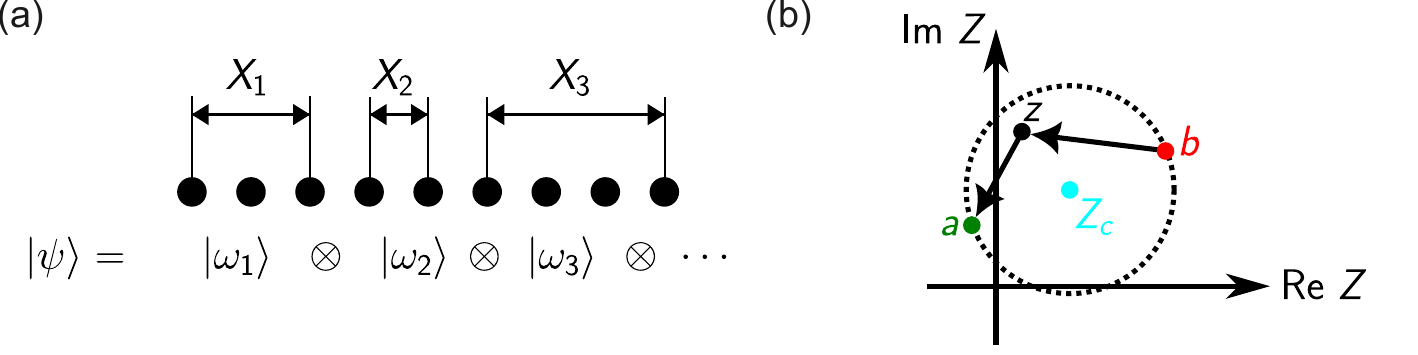}
    \caption{(a) The definition of entanglement depth. 
    A wave function $|\psi\rangle$ has an entanglement depth of $m$ if it can be decompose into tensor product of inseparable states, each containing at most $m$ sites.
    (b) The possible values of a complex random variable $Z$ distributed on the dashed circle on the complex plane.
    Points $a$ and $b$ are the two ends of one diameter, and $Z_c = (a + b)/2$ is the center of the circle.
    }
    \label{fig:depth}
\end{figure}

For a pure state $|\psi\rangle$, this QFI is related to the equal-time correlation function of the operator $\widehat{\cal O}$, i.e.,
\begin{equation}
    F_Q(|\psi\rangle, \widehat{\cal O}) = 4\var (\widehat{\mathcal O} ) \equiv 4 \left( \langle \widehat{\mathcal O}^2 \rangle - \langle \widehat{\mathcal O} \rangle^2 \right) \, .
\end{equation}
The QFI is related the entanglement of $|\psi\rangle$.
Consider a quantum system defined on a lattice with $N_L$ degrees of freedom. 
If it is ``disentangled'' into multiple ``patches'' [see Fig.~\ref{fig:depth}(a)], its wave function will have the following form:
\begin{equation}\label{eqn:def-m-entangled}
    |\psi \rangle = \bigotimes_{j}|\omega_j\rangle\,,
\end{equation}
in which $|\omega_j\rangle$ is the state of the $j$-th ``patch'' of the system whose support is denoted as $X_j$. 
The entanglement depth, can be defined as the maximum value of the sizes of these patches:
\begin{equation}
    m = \max_{j} \{|X_j|\}\,.
\end{equation}
One can prove that the QFI density is upper bounded by:
\begin{equation}\label{eqn:qfi-upper-bound}
    f_Q \equiv \frac{F_Q(|\psi\rangle, \widehat{\cal O})}{N_L} \leq m (h_{\text{max}}-h_{\text{min}})^2\,.
\end{equation}
Here $h_{\rm max}$ and $h_{\rm min}$ stand for the maximum and minimum eigenvalues of the local operators $\widehat{\cal O}$.
If the QFI is larger than the bound, the system is at least $(m+1)$-partite entangled. 
Note that the QFI depends on the choice of the operator $\widehat{\cal O}$, and the bound is a sufficient condition for the entanglement detection.
Therefore, it serves as an entanglement witness in a quantum system.

The discussion in the previous paragraphs is based on the assumption that all the local operators in $\widehat{\cal O}$  are hermitian.
In some context, the quantum Fisher information can also be defined when the operators are not hermitian.
We will prove that a similar entanglement upper bound still holds for a modified definition of QFI, given by:
\begin{equation}
    F_Q(|\psi\rangle, \widehat{\cal O}) = 4 \langle \widehat{\cal O}^\dagger \widehat{\cal O}\rangle - 4 |\langle \widehat{\cal O} \rangle|^2\,, 
\end{equation}
when the operator $\widehat{\cal O}$ takes the following form:
\begin{equation}
    \widehat{\cal O} = \sum_{j} \widehat{O}_{X_j} = \sum_{j}\sum_{i \in X_j} \widehat{\cal O}_{i} e^{i\chi_i}\,.
\end{equation}
Here the operators defined on each {\it site} $\widehat{\cal O}_i$ are still Hermitian, and the phase factors $\chi_i \in \mathbb{R}$.
This form is applicable to a wide class of operators and naturally extends 
the previous framework for using the QFI as entanglement witness.
For example, spin density operators with a finite momentum $\boldsymbol{q}$ have phase factors $\chi_i = \boldsymbol{q}\cdot \boldsymbol{r}_i$ in their definition.
Due to these phase factors, the operators defined on each {\it patch} $\widehat{\cal O}_{X_j}$ can be non-hermitian.

In this case, we can show that the quantum Fisher information $F_Q$ has the following form:
\begin{align}
    F_Q(|\psi\rangle, \widehat{\cal O}) =& 4\sum_{jj'}\langle \psi| \widehat{\cal O}^\dagger_{X_j} \widehat{\cal O}_{X_{j'}} |\psi \rangle - 4\left|\sum_j \langle\psi| \widehat{\cal O}_{X_j} | \psi \rangle\right|^2\nonumber\\
    =&4\sum_j \langle \omega_j| \widehat{\cal O}^\dagger_{X_j} \widehat{\cal O}_{X_j} | \omega_j \rangle + 4\sum_{j\neq j'}\ \langle \omega_j| \widehat{\cal O}^\dagger_{X_j} | \omega_{j} \rangle \langle \omega_{j'}| \widehat{\cal O}_{X_{j'}} | \omega_{j'} \rangle  - 4\sum_{jj'}\langle \omega_j| \widehat{\cal O}^\dagger_{X_j} | \omega_{j} \rangle \langle \omega_{j'}| \widehat{\cal O}_{X_{j'}} | \omega_{j'} \rangle\nonumber\\
    =& 4\sum_j \langle \omega_j| \widehat{\cal O}_{X_j}^\dagger \widehat{\cal O}_{X_j} | \omega_j \rangle - 4\sum_{j}\ \left|\langle \omega_j| \widehat{\cal O}_{X_j} | \omega_{j} \rangle \right|^2\,,\label{eqn:qfi-complex-variance}
\end{align}
which is the summation of the ``variance'' of the operators $\widehat{\cal O}_{X_j}$ defined in each patch. 
We note that although the operators $\widehat{\cal O}_{X_j}$ are no longer Hermitian, they are still ``normal'' matrices, given that they commute with their own Hermitian conjugate.
Therefore, eigenvalues and eigenstates can still be defined for $\widehat{\cal O}_{X_j}$ operators.
By assuming the eigenvalues of the local operators on each {\it site} are $h_\alpha \in \mathbb{R}$, the eigenvalues of the patch operator $\widehat{\cal O}_{X_j}$ can be labeled by a collection of single-site eigenvalue indices $\{\alpha_i\}$:
\begin{equation}
    Z_{\{\alpha_i\}} = \sum_{i\in X_j} h_{\alpha_i} e^{i\chi_i}\,.
\end{equation}
As such, each term in Eq.~(\ref{eqn:qfi-complex-variance}) can be considered as a generalized ``variance'' of a complex random variable:
\begin{equation}\label{eqn:patch-complex-qfi-variance-form}
     \langle \omega_j| \widehat{\cal O}_{X_j}^\dagger \widehat{\cal O}_{X_j} | \omega_j \rangle - |\langle \omega_j| \widehat{\cal O}_{X_j} | \omega_j \rangle|^2 = \sum_{\{\alpha_i\}} p_{\{ \alpha_i\}} Z^*_{\{\alpha_i\}} Z_{\{\alpha_i\}} - \left|\sum_{\{\alpha_i\}}p_{\{ \alpha_i\}} Z_{\{\alpha_i\}}\right|^2\,,
\end{equation}
in which $p_{\{\alpha_i\}} = |\langle \{\alpha_i\}| \omega_j\rangle|^2$ is the probability of the $\widehat{\cal O}_{X_j}$ eigenstate $|\{\alpha_i\}\rangle$ in the state $|\omega_j\rangle$.

We then study the properties of the distribution of $\widehat{\cal O}_{X_j}$ eigenvalues. 
Since $h_{\rm min} \leq h_{\alpha} \leq h_{\rm max}$, 
we can define the ``center'' of the distribution as $h_c = (h_{\rm min} + h_{\rm max})/2$. 
Therefore, for any local eigenvalue $h_\alpha$, 
it will always satisfy $|h_\alpha - h_c| < (h_{\rm max} - h_{\rm min})/2$.
Using $h_\alpha = (h_\alpha - h_c) + h_c$ to rewrite Eq.~(\ref{eqn:patch-complex-qfi-variance-form}), we end up with the following equation:  
\begin{equation}
    Z_{\{\alpha_i\}} = \sum_{i\in X_j} (h_{\alpha_i} - h_c) e^{i\chi_i} + h_c\sum_{i\in X_j}e^{i\chi_i}\,.
\end{equation}
The second term is irrelevant to the choice of $\{\alpha_i\}$, which can be considered as the ``center'' of the distribution of $\widehat{\cal O}_{X_j}$ eigenvalues. For convenience, we define this point as follows:
\begin{equation}
    Z_c = h_c\sum_{i\in X_j}e^{i\chi_i}\,.
\end{equation}
One can further show that all $\widehat{\cal O}_{X_j}$ eigenvalues distribute within a circle centered around $Z_c$, as illustrated in Fig.~\ref{fig:depth}(b).
Using triangle inequality, the upper bound of the radius can be shown as:
\begin{equation}
    \left|Z_{\{\alpha_i\}} - Z_c\right| \leq \sum_{i\in X_j} |h_{\alpha_i} - h_c| \leq l_j \frac{h_{\rm max} - h_{\rm min}}{2}\,.
\end{equation}
Here we have defined the size of patch $X_j$ as $l_j$.

To prove the upper bound of Eq.~(\ref{eqn:patch-complex-qfi-variance-form}), we first choose two points $a$ and $b$ along an arbitrary diameter of this circle, which satisfies $Z_c = (a+b)/2$ and $|a - b| = l_j (h_{\rm max} - h_{\rm min})$.
Since all $Z_{\{\alpha_i\}}$ values are within this circle, we will always have the following inequality:
\begin{equation}
    {\rm Re}\,\left[(a - Z_{\{\alpha_i\}})(Z_{\{\alpha_i\}}^* - b^*)\right] \geq 0\,.
\end{equation}
This is because of the angle between the two arrows shown in Fig.~\ref{fig:depth}(b) is always smaller than $90^\circ$ if $z$ is within the dashed circle.
Such an inequality can be easily transformed into the following form:
\begin{equation}
    Z^*_{\{\alpha_i\}}Z_{\{\alpha_i\}} \leq {\rm Re}\,\left[(a+b)Z^*_{\{\alpha_i\}} - ab^*\right]\,.
\end{equation}
Clearly, it leads to an upper bound for Eq.~(\ref{eqn:patch-complex-qfi-variance-form}):
\begin{align}
      \langle \omega_j| \widehat{\cal O}_{X_j}^\dagger \widehat{\cal O}_{X_j} | \omega_j \rangle - |\langle \omega_j| \widehat{\cal O}_{X_j} | \omega_j \rangle|^2 &\leq \sum_{\{\alpha_i\}} p_{\{\alpha_i\}} {\rm Re}\,\left[(a+b)Z^*_{\{\alpha_i\}} - ab^*\right] - \left|\sum_{\{\alpha_i\}}  p_{\{\alpha_i\}} Z_{\{\alpha_i\}} \right|^2 \nonumber \\
      &= {\rm Re}\,\left[(a+b)\langle \widehat{\cal O}_{X_j}\rangle^* -ab^*\right] - |\langle \widehat{\cal O}_{X_j} \rangle|^2 \nonumber\\
      &= |Z_c|^2 - |\langle \widehat{\cal O}_{X_j}\rangle - Z_c|^2 - {\rm Re}\,ab^* \nonumber \\
      & \leq |Z_c|^2 - {\rm Re}\,ab^* \nonumber \\ 
      & = \frac{1}{4} |a - b|^2 \nonumber \\ 
      & = l_j^2 \frac{(h_{\rm max} - h_{\rm min})^2}{4}\,.
\end{align}
We note that this upper bound shares the same format as the one occurring in the case with Hermitian $\widehat{\cal O}_{X_j}$ operators.
Indeed, this is a generalization of {\it Popoviciu's inequality} and {\it Bhatia–Davis inequality} \cite{Bhatia2000Better} into the case with complex random variables.

By summing over all patches, the QFI for the entire system is therefore upper bounded by:
\begin{equation}
    F_Q(|\psi\rangle, \widehat{\cal O}) \leq (h_{\rm max} - h_{\rm min})^2\sum_j l_j^2\,.
\end{equation}
The patch sizes $l_j$ satisfy $\sum_j l_j = N_L$ and $0 \leq l_j \leq m$, in which $N_L$ is the total amount of sites in the system.
For any two positive integers $<m$ with fixed sum, the maximum value of the sum of their squares is achieved when one of them is as large as possible.
We can then iteratively apply this argument to the patch sizes $l_j$ to show that the maximum value of the sum $\sum_j l_j^2$ is achieved when there are $\lfloor \frac{N_L}{m} \rfloor$ patches with size $m$ and one extra patch with size $r = N - m\lfloor \frac{N_L}{m} \rfloor < m$.
Hence, the upper bound of the QFI is:
\begin{align}
    F_Q(|\psi\rangle, \widehat{\cal O}) \leq & (h_{\rm max} - h_{\rm min})^2 \left( m^2 \lfloor \frac{N_L}{m} \rfloor + r^2 \right)\nonumber\\
    \leq & (h_{\rm max} - h_{\rm min})^2 \left( m^2 \lfloor \frac{N_L}{m} \rfloor + rm \right)\nonumber\\
    = & (h_{\rm max} - h_{\rm min})^2 m \left(m \lfloor \frac{N_L}{m}\rfloor  + r\right) \nonumber\\
    = & (h_{\rm max} - h_{\rm min})^2 m N_L\,.
\end{align}
This upper bound of the modified QFI is still the same as Eq.~(\ref{eqn:qfi-upper-bound}).
Therefore, we conclude that non-Hermitian operators, such as the spin density operators at an incommensurate $\boldsymbol{q}$ can still be used to detect the entanglement in a quantum state.
We also note that the proof of this bound only requires the local operators commute~\cite{Fang2024Amplified} while locality is not used. Therefore, commuting nonlocal operators also fit this framework.

\section{Equivalence between the nQFI of the local spin and charge density operators}

In the main text we show the nQFI of local spin operator is always trivial for noninteracting fermions. 
Here we show the bound for local charge operator  is equal to the bound for local spin operator.

When the spin $SU(2)$ symmetry is present, it forces $\langle n_{\uparrow i} \rangle = \langle n_{\downarrow i} \rangle$.
Then the nQFI of the local spin operator can be expressed as
\begin{align}
    {\rm nQFI} &= \frac{1}{N_L}\sum_{ij} e^{i{\bm q}\cdot({\bm r}_{i}-{\bm r}_{j})} \langle (n_{\uparrow i}-n_{\downarrow i}) (n_{\uparrow j}-n_{\downarrow j}) \rangle  \n \\
    &= \frac{1}{N_L}\sum_{ij} e^{i{\bm q}\cdot({\bm r}_{i}-{\bm r}_{j})}  \Big( 
    \langle (n_{\uparrow i}+n_{\downarrow i}) (n_{\uparrow j}+n_{\downarrow j}) \rangle - 2\langle n_{\downarrow i} \rangle \langle n_{\uparrow j} \rangle - 2\langle n_{\uparrow i} \rangle \langle n_{\downarrow j} \rangle \Big) \nonumber\\ 
    &= \frac{1}{N_L}\sum_{ij} e^{i{\bm q}\cdot({\bm r}_{i}-{\bm r}_{j})}  \Big( \langle n_{i}n_{j} \rangle - \langle n_{i}\rangle \langle n_{j} \rangle\Big) 
\end{align}
where $n_{i} = n_{\uparrow i}+n_{\downarrow i}$. 
Note the difference in the normalization factors that are considered: the charge operator has $h_{\text{max}}=1$, $h_{\text{min}}=0$ while the spin operator has $h_{\text{max}}=1/2$, $h_{\text{min}}=-1/2$.
Thus, for noninteracting electrons, the nQFI of the local spin operator is the same as the nQFI of the local charge operator. 
Therefore, the upper bound we proved for the local spin nQFI also applies to its charge counterpart.

\section{Upper bound of QFI with local operator for noninteracting multiorbital electron system}

We define general density operators $\widehat{\cal O}(\bm{q})=\sum_{i\alpha}e^{i\bm{q}\cdot \bm{r}_i}\widehat{\cal O}_{i\alpha}$ at wave vector $\bm{q}$, with $\widehat{\cal O}_{i\alpha}=\frac{1}{2}\sum_{\sigma\sigma'}c^{\dagger}_{i\alpha\sigma}[\sigma^{a}]_{\sigma\sigma'}c_{i\alpha\sigma}$ for each orbital $\alpha$ and each site $i$, where $\alpha=1,\cdots, M$, $a = 0$ or $z$ represents charge or spin channel respectively. The $\widehat{\cal O}(\bm{q})=\sum_{i\alpha}\widehat{\cal O}_{i\alpha}(\bm{q})$ in the band representation has the form:
\begin{align}
   \widehat{\cal O}(\bm{q})=\frac{1}{2}\sum_{\bm{k}mn}c^{\dagger}_{\bm{k}+\bm{q},m\sigma}[\sigma^{a}]_{\sigma\sigma'}K_{mn}(\bm{k},\bm{q})c_{\bm{k},n\sigma'} \, ,
\end{align}
where $\sigma,\sigma'= \uparrow,\downarrow$ are spin labels, $m,n = 1,\cdots,M$ are band labels, $K_{mn}(\bm{k},\bm{q})=\sum_{\alpha}u^{*}_{\alpha m}(\bm{k}+\bm{q})u_{\alpha n}(\bm{k})$ where $u_{\alpha n}(\bm{k})$ is the Bloch function that transforms electrons from orbital basis to band basis. In the band representation, the multiorbital Hamiltonian is exactly diagonalized:
\begin{align}
    H=\sum_{\bm{k}m\sigma} \epsilon_{\bm{k}m}c^{\dagger}_{\bm{k}m\sigma}c_{\bm{k}m\sigma}\,,
\end{align}
where $\epsilon_{\bm{k}m}$ is the dispersion for each band $m$. These bands could be non-degenerate.
The QFI for the general density operator is:
\begin{align}
    f_{Q}({\bm{q}})=&\frac{4}{MN_{L}}[\langle \widehat{\cal O}^{\dagger}(\bm{q})\widehat{\cal O}(\bm{q})\rangle-\langle \widehat{\cal O}^{\dagger}(\bm{q})\rangle \langle \widehat{\cal O}(\bm{q})\rangle]\\
    =&\frac{1}{MN_{L}}\sum_{\bm{k}\bm{k}'mnm'n'}\sum_{\sigma\sigma'\mu\mu'=\uparrow\downarrow}\langle c^{\dagger}_{\bm{k}n\sigma'}c_{\bm{k}',n'\mu'}\rangle\langle c_{\bm{k}+\bm{q},m'\sigma}c^{\dagger}_{\bm{k}'+\bm{q},m\mu}\rangle K^{*}_{mn}(\bm{k},\bm{q})K_{m'n'}(\bm{k}',\bm{q}) [\sigma^{a}]_{\sigma\sigma'}[\sigma^{a}]_{\mu\mu'}\\
    =&\frac{2}{MN_{L}}\sum_{\bm{k}mn}K^{*}_{mn}(\bm{k},\bm{q})K_{mn}(\bm{k},\bm{q})f({\epsilon_{\bm{k},n}}) [1-f(\epsilon_{\bm{k}+\bm{q}, m})]\,,
\end{align}
where $f({\epsilon_{\bm{k} n}})=\theta(-{\epsilon_{\bm{k} n}})$ is the Fermi-Dirac distribution for band $n$ at zero temperature.  Same as the one band version, the QFI density is upper-bounded  through the arithmetic mean-geometric mean inequality:
\begin{align}
    f_{Q}(\bm{q})\leq &\frac{1}{MN_{L}}\sum_{\bm{k}mn}K^{*}_{mn}(\bm{k},\bm{q})K_{mn}(\bm{k},\bm{q})[f({\epsilon_{\bm{k} ,n}})^{2} + (1-f(\epsilon_{\bm{k}, m}))^2]\\
    =&\frac{1}{MN_{L}}\sum_{\bm{k}mn}K^{*}_{mn}(\bm{k},\bm{q})K_{mn}(\bm{k},\bm{q})[f({\epsilon_{\bm{k},n}}) + 1-f(\epsilon_{\bm{k}+\bm{q,} m})]\\
    =&\frac{1}{MN_{L}}\sum_{\bm{k}}{\rm Tr}(K^{\dagger}(\bm{k},\bm{q})K(\bm{k},\bm{q}))\\
    =&1\,,
\end{align}
where in the last two equations we utilized the unitary matrix property of $K(\bm{k},\bm{q})$: $\sum_{m}K^{*}_{mn}(\bm{k},\bm{q})K_{mn}(\bm{k},\bm{q})=\sum_{m\alpha\beta}u^{*}_{\alpha n}(\bm{k})u_{\alpha m}(\bm{k}+\bm{q})u^{*}_{\beta m}(\bm{k}+\bm{q})u_{\beta n}(\bm{k})=\sum_{\alpha}u^{*}_{\alpha n}(\bm{k})u_{\alpha n}(\bm{k})=1$. The first inequility saturates when ${\rm sgn}(\epsilon_{\bm{k},m})=-{\rm sgn}(\epsilon_{\bm{k}+\bm{q},n})$ for any momentum $\bm{k}$ and for any two bands $m,n$.

\section{Upper bound of the QFI with local operators for a superconductor}

In this section we study the QFI of the superconducting BdG states and the charge operator $\widehat{\cal O}_i=e^{i {\bm q} \cdot {\bm r}_i}n_i$.
Here we label all the modes, including site and spin, by $i$.
A general BdG ground state is annihilated by the operators
\begin{align}
    \label{eqn:annih_singular}
    \gamma_n = \sum_{i=1}^{N_L} c_i U^*_{in} + c^\dagger_i V^*_{in}\,,
\end{align}
where $n=1\dots,N_L$, $U$ and $V$ are square matrices that appear in the eigenstates of the BdG Hamiltonian:
\begin{equation}
    H_{BdG} \Psi = \Psi 
    \begin{pmatrix}
        D &  \\
          & -D
    \end{pmatrix}\,, \qquad 
    \Psi = \begin{pmatrix}
        U & V^* \\
        V & U^*
    \end{pmatrix}
\end{equation}
where $D$ is a diagonal matrix with positive eigen-energies. 
These matrices satisfy the conditions
\begin{align}
    U^\dagger U + V^\dagger V & = UU^\dagger + V^*V^T = \mathbb{1}\,,\label{eqn:bdg-uv-orthonormal-1}\\
    U^{\dagger} V^* + V^\dagger U^* & = UV^\dagger + V^*U^T = \mathbb{0}\,.\label{eqn:bdg-uv-orthonormal-2}
\end{align}
The correlation functions are:
{\begin{align}
    \langle c^\dagger_i c_j\rangle &= (VV^\dagger)_{ij}\equiv G_{ij}\,, \\
    \langle c^\dagger_i c^\dagger_j\rangle &= (VU^\dagger)_{ij} \equiv F_{ij}\,.
\end{align}

Therefore, the nQFI of charge operator can be expressed as
\begin{align}
    {\rm nQFI} &= \frac{4}{N_L}\sum_{ij}e^{i{\bm q}\cdot({\bm r}_{i}-{\bm r}_{j})} \Big( \langle c^\dagger_i c_i c^\dagger_j c_j\rangle - \langle c^\dagger_i c_i\rangle\langle c^\dagger_j c_j\rangle \Big) \n\\
    &= \frac{-4}{N_L}\sum_{ij}e^{i{\bm q}\cdot({\bm r}_{i}-{\bm r}_{j})} \Big( \langle c^\dagger_i c_j\rangle \langle c^\dagger_j c_i\rangle + \langle c^\dagger_i c^\dagger_j\rangle \langle c_i c_j\rangle \Big) \n\\
    &=\frac{4}{N_L}\sum_{ij}e^{i{\bm q}\cdot({\bm r}_{i}-{\bm r}_{j})} (-G_{ij}G_{ji} + |F_{ij}|^2) \n\\
    &\leq \frac{4}{N_L} \Tr (V^\dagger\Sigma V U^\dagger\Sigma^\dagger U) \n\\
    &\leq 2\,,
\end{align}
where $[\Sigma]_{ij}=e^{i{\bm q}\cdot{\bm r}_{i}} \delta_{ij}$ and in the fifth line we used Cauchy-Schwartz inequality.

For the local spin operators, the nQFI bound can be derived in a similar manner
\begin{align}
    {\rm nQFI} &= \frac{1}{N_L}\sum_{i,j=1}^{N_L}e^{i{\bm q}\cdot({\bm r}_{i}-{\bm r}_{j})} \Big( \langle (c^\dagger_{i\uparrow} c_{i\uparrow}-c^\dagger_{i\downarrow} c_{i\downarrow}) (c^\dagger_{j\uparrow} c_{j\uparrow}-c^\dagger_{j\downarrow} c_{j\downarrow})\rangle - \langle c^\dagger_{i\uparrow} c_{i\uparrow}-c^\dagger_{i\downarrow} c_{i\downarrow}\rangle^2\Big) \n\\
    &=\frac{1}{N_L}\sum_{a\neq b=1}^{2N_L}e^{i{\bm q}\cdot({\bm r}_{a}-{\bm r}_{b})+i\pi(a+b)} \langle n_a n_b \rangle \n \\
    &\leq \frac{1}{N_L} \Tr (V^\dagger\Sigma V U^\dagger\Sigma^\dagger U) \n\\
    &\leq 1\,,
\end{align}
where $n_{2a} = c^\dagger_{a\uparrow} c_{a\uparrow}$, $n_{2a+1} = c^\dagger_{a\downarrow} c_{a\downarrow}$ and $[\Sigma]_{ab}=e^{i{\bm q}\cdot{\bm r}_{a}+i\pi a} \delta_{ab}$.

\section{Interaction-Induced Enhancement of the QFI in Fermi Liquids}

As is shown in the main text, the quantum Fisher information for noninteracting electrons has its upper bound - one. Next we show that the QFI for interacting electrons at RPA level is larger than noninteracting counterparts, but the magnitude is still trivial.

We aim to prove the inequality
\begin{align}
    \frac{\chi^{\prime\prime}_{RPA}}{\chi^{\prime\prime}_{0}}=\frac{1}{(1-U\,(\chi^{\prime}_{0}))^2+(U\chi^{\prime\prime}_{0})^2}\geq 1\,,
\end{align}
where $\chi^{\prime}\equiv {\rm Re}\chi$ and $\chi^{\prime\prime}\equiv {\rm Im}\chi$ are real and imaginary part of spin susceptibility, respectively.
This inequality is equivalent to
\begin{align}
    U^{2}[( \chi_{0}^{\prime})^2+( \chi_{0}^{\prime\prime})^2 ]\leq 2U  \chi_{0}^{\prime},,
\end{align}  
where the noninteracting spin susceptibility $\chi_{0}(\bm{q},\omega)$ has the following form 
\begin{align}
    \chi_{0}(\bm{q},\nu-i\delta)
  =\frac{2}{N_{L}}\sum_{\bm{k}}\frac{f(\epsilon_{\bm{k}+\bm{q}})-f(\epsilon_{\bm{k}})}{\nu-(\epsilon_{\bm{k}+\bm{q}}-\epsilon_{\bm{k}})-i\delta}\,.
\end{align}
To establish this result, we consider the following observations:  
\begin{itemize}
    \item Due to the relation \( x\delta(x) = 0 \) and hence \( 1/x \gg \pi\delta(x) \), it follows that  
   \begin{align}
       |\chi_{0}^{\prime}| \gg | \chi_{0}^{\prime\prime}|.
   \end{align}  
   Consequently, we obtain the bound  
   \begin{align}
       U^{2}[(\chi_{0}^{\prime})^2+(\chi_{0}^{\prime\prime})^2 ] \leq 2U^{2} (\chi_{0}^{\prime})^2.
   \end{align}  

    \item In a Fermi liquid, the Stoner criterion imposes the condition for the stability condition of the Fermi liquid
   \begin{align}
       1 - U  \chi_{0}^{\prime}(\bm{q},\omega) \geq 0.
   \end{align}  
   This implies that  
   \begin{align}
       U^{2}(\chi_{0}^{\prime})^2 \leq U \chi_{0}^{\prime}.
   \end{align}  
   Combining this with the previous bound, we conclude that  
   \begin{align}
       U^{2}[(\chi_{0}^{\prime})^2+(\chi_{0}^{\prime\prime})^2 ] \leq 2U^{2} (\chi_{0}^{\prime})^2 \leq 2U \chi_{0}^{\prime},
   \end{align} 
\end{itemize}
   thereby proving the desired inequality.

\section{Correlation function of the Jordan-Wigner transformed operators}
The correlation functions of Jordan-Wigner transformation string operators in one dimension can be solved numerically using a Toeplitz matrix~\cite{PFEUTY197079,Barouch1971Statistical,sachdev1999quantum,Mbeng2024Ising}.
The correlation function between the sites $j_1$ and $j_2$ is
\begin{equation}\label{eqn:corr_TFIM}
    4\langle S^x_{j_1}S^x_{j_2} \rangle = \det 
    \begin{pmatrix}
        M_{j_1,j_1+1} & M_{j_1,j_1+2} & \cdots & M_{j_1,j_2-1} & M_{j_1,j_2} \\
        M_{j_1+1,j_1+1} & M_{j_1+1,j_1+2} & \cdots & M_{j_1+1,j_2-1} & M_{j_1+1,j_2} \\
        \vdots & \vdots & \ddots & \vdots & \vdots \\
        M_{j_2-2,j_1+1} & M_{j_2-2,j_1+2} & \cdots & M_{j_2-2,j_2-1} & M_{j_2-2,j_2} \\
        M_{j_2-1,j_1+1} & M_{j_2-1,j_1+2} & \cdots & M_{j_2-1,j_2-1} & M_{j_2-1,j_2}
    \end{pmatrix}
\end{equation}
where the spin operator is $2S^x_j=a_j+a_j^\dagger$,
\begin{align}
    M_{jj'} &= \delta_{jj'} - 2(G_{jj'}+F_{jj'})
\end{align}
For a BdG state:
\begin{align}
    G_{jj'} &= \langle c_j c^\dagger_{j'} \rangle = \frac 1L \sum_k e^{ik(j-j')}|u_k|^2 \\
    F_{jj'} &= \langle c_j c_{j'} \rangle = -\frac 1L \sum_k e^{ik(j-j')} u_k^* v_k 
\end{align}
and $u_k$ and $v_k$ are the two components of eigenstates of the BdG Hamiltonian.
For a noninteracting fermion state:
\begin{align}
    G_{jj'} &= \langle c_j c^\dagger_{j'} \rangle = \frac 1L \sum_{|k|\leq k_F} e^{ik(j-j')} \\
    F_{jj'} &= 0
\end{align}
where fermoins are filled at momenta $|k|\leq k_F$. 

Then the nQFI of the Jordan Wigner transformed operator at the wave vector $q$ is:
\begin{equation}
    {\rm nQFI} = \frac{1}{N_L} \sum_{i,j=1}^{N_L} 4\langle S^x_{j_1}S^x_{j_2} \rangle e^{iq(j_1-j_2)}
\end{equation}

For a ladder model with extended string operator $\theta^{\rm sign}_{ij}$, one can always stretch the ladder into a one dimensional chain and arrange the sites in an order that is consistent with the extended string operator. The correlation functions can then be evaluated in the same way as the one dimensional case.


\end{document}